\documentclass[a4paper,11pt]{article}

\usepackage{hyperref}%
\hypersetup{colorlinks=true,%
  linkcolor=blue,%
  citecolor=magenta,%
  urlcolor=black,%
  bookmarksnumbered=true,%
  bookmarkstype=toc,%
  bookmarksopen=false,%
  pdftitle={Models for the BPS Berry Connection},%
  pdfkeywords={Berry connection; BPS monopole; supersymmetric quantum
    mechanics},%
  pdfauthor={Satoshi Ohya}}%

\usepackage[margin=1in]{geometry}%

\usepackage[sb]{libertine}%
\usepackage[T1]{fontenc}%
\usepackage{textcomp}%
\usepackage[varqu,varl]{zi4}%
\usepackage{bm}%
\usepackage{libertinust1math}%
\usepackage[cal=boondoxo,scr=boondoxo,bb=boondox]{mathalfa}%
\allowdisplaybreaks%

\usepackage{titlesec}%
\titleformat{\section}[block]{\filright\bfseries}{\thesection.}{0.5em}{}[]%
\titleformat{\subsection}[block]{\filright\bfseries}{\thesubsection.}{0.5em}{}%

\usepackage{enumitem}%

\usepackage{cite}%

\title{\large\bfseries Models for the BPS Berry Connection}
\author{
  {\normalsize Satoshi Ohya}\\[1em]
  {\small\textit{Institute of Quantum Science, Nihon University}}\\
  {\small\textit{Kanda-Surugadai 1-8-14, Chiyoda, Tokyo 101-8308,
      Japan}}\\[1ex]
  {\small\texttt{ohya.satoshi@nihon-u.ac.jp}}}
\date{\small (Dated: \today)}

\begin{document}
\maketitle%
\flushbottom%

\begin{abstract}
  Motivated by the Nahm's construction, in this paper we present a
  systematic construction of Schr\"{o}dinger Hamiltonians for a
  spin-1/2 particle where the Berry connection in the ground-state
  sector becomes the Bogomolny-Prasad-Sommerfield (BPS) monopole of
  $SU(2)$ Yang-Mills-Higgs theory. Our construction enjoys a single
  arbitrary monotonic function, thereby creating infinitely many
  quantum-mechanical models that simulate the BPS monopole in the
  space of model parameters.
\end{abstract}

\section{Introduction}
\label{section:1}
The Bogomolny-Prasad-Sommerfield (BPS) monopole
\cite{Prasad:1975kr,Bogomolny:1975de} is the simplest yet most
profound example for non-Abelian magnetic monopoles. Originally, it
just appeared as the simplest analytic expression for the 't
Hooft-Polyakov monopole \cite{'tHooft:1974qc,Polyakov:1974ek} of
$SU(2)$ Yang-Mills-Higgs theory by taking the limit of vanishing Higgs
potential. However, it was soon recognized that the BPS monopole has
an amazingly rich mathematical structure. In particular, it was
realized that the BPS monopole can be constructed without solving the
field equations. To date, there exist several constructive approaches
to the BPS monopoles, the most notable of which is the Nahm's
construction \cite{Nahm:1979yw}. As is well-known, this approach
consists of the following three steps (see, for example, Section 4.4.1
of Ref.~\cite{Weinberg:2006rq}):
\begin{enumerate}[label=(\arabic*),leftmargin=1.5em]
\item Solve the Nahm equation---the first-order nonlinear matrix
  differential equation with quadratic nonlinearity---and obtain the
  Nahm data;
\item Solve the construction equation---a one-dimensional Dirac-like
  equation defined through the Nahm data; and
\item Compute the following:
  \begin{align}
    A^{i}_{ab}(\bm{x})=i\langle\Psi_{a}(\bm{x})|\frac{\partial}{\partial x_{i}}|\Psi_{b}(\bm{x})\rangle,\label{eq:1}
  \end{align}
  where $|\Psi_{a}(\bm{x})\rangle$ stand for the normalizable
  solutions of the construction equation.
\end{enumerate}
It can then be shown that Eq.~\eqref{eq:1} satisfies the Bogomolny
equation---the defining equation of the BPS monopole---and indeed
describes the BPS monopole.

Now, an important observation here is that Eq.~\eqref{eq:1} takes
exactly the same form as the non-Abelian Berry connection
\cite{Wilczek:1984dh}. This implies that, if
$|\Psi_{a}(\bm{x})\rangle$ are realized as wavefunctions for a
degenerate energy level, the BPS monopoles may well be simulated as
the Berry connections in the parameter space of ordinary
quantum-mechanical systems. In fact, such systems do exist and so far
there have been discovered two examples. The first example is given in
Ref.~\cite{Sonner:2008be}, where it has been discussed that a spin-1/2
particle on $\mathbb{S}^{2}$ with specific magnetic field and
potential enjoys a non-Abelian Berry phase described by the BPS
monopole. The second example is given in Ref.~\cite{Ohya:2015xya},
where the author has shown that a free spinless particle on
$\mathbb{S}^{1}$ with particular pointlike interactions can be
effectively described by a spin-1/2 particle on an interval and yields
the Berry connection that describes the BPS monopole. A natural
question that arises is then whether there exist any other models that
reproduce the BPS monopole. As we shall see in the rest of the paper,
the answer to this question is affirmative: there exist infinitely
many nonrelativistic quantum-mechanical systems where the Berry
connection in the ground-state sector becomes the BPS monopole of
four-dimensional $SU(2)$ Yang-Mills-Higgs theory. The goal of this
short paper is to show this and present several new examples.

The rest of the paper is organized as follows. In Section
\ref{section:2} we first introduce two distinct two-component
wavefunctions in one dimension, both of which are nodeless, mutually
orthogonal, and specified by a single monotonically increasing
function $W$. In terms of the Nahm's construction, these wavefunctions
correspond to the solutions of the construction equation. We then show
that the non-Abelian Berry connection built upon these wavefunctions
is nothing but the BPS monopole of $SU(2)$ Yang-Mills-Higgs theory. It
is also shown that the matrix elements of $W$ generally becomes the
BPS solution for the Higgs field. In Section \ref{section:3} we
construct a family of one-dimensional quantum-mechanical models for a
spin-1/2 particle by using the technique of supersymmetric quantum
mechanics. In this family the ground states are doubly degenerate and
the ground-state wavefunctions are given by those constructed in
Section \ref{section:2}. The non-Abelian Berry connection in the
ground-state sector is therefore always given by the BPS monopole. We
also discuss that our models enjoy an exotic supersymmetry called the
second-order derivative supersymmetry
\cite{Andrianov:1993md,Andrianov:1994aj}. Section \ref{section:4}
presents several examples to illustrate our construction. We shall see
that our construction method yields all the existing
models\footnote{The BPS Berry connection has also been discussed in
  the context of topological insulators \cite{Hashimoto:2016dtm},
  where the solutions of the construction equation are realized as the
  solutions of four-dimensional Dirac equation with particular
  boundary conditions. Note that in this paper we will focus on
  nonrelativistic quantum mechanics and not touch upon the Dirac
  equation.} as well as new ones.

For the sake of notational brevity, throughout the paper we will work
in arbitrary dimensionless units, which can always be converted into
the physical units by appropriate rescaling.

\section{BPS Berry connection}
\label{section:2}
In the standard approach to quantum mechanics, one first constructs a
Hamiltonian and then solves the Schr\"{o}dinger equation. In this
paper, however, we solve the problem in reverse order; that is, we
first start with a desired ground-state wavefunction and then
construct a Hamiltonian. This is possible because, as is well-known
especially in the context of supersymmetric quantum mechanics, the
ground-state wavefunction generally determines potential energy. In
this Section we shall first introduce two nodeless wavefunctions,
which correspond to the solutions of the construction equation. Then
we shall show that the BPS monopole and Higgs solutions of $SU(2)$
Yang-Mills-Higgs theory are respectively given by the Berry connection
and some matrix elements with respect to these wavefunctions. In the
subsequent Section we shall construct a family of Schr\"{o}dinger
Hamiltonians whose lowest-energy eigenstates are given by the
wavefunctions constructed in this Section.

To begin with, let $I\subseteq\mathbb{R}$ be a one-dimensional
subspace, which can be either a finite interval or an infinite
interval, and $z$ be the coordinate of $I$ with $z_{+}$ and
$z_{-}(<z_{+})$ being two endpoints of $I$. Let us then consider the
following two wavefunctions on $I$:
\begin{align}
  \psi_{\pm}(z)=N\sqrt{W^{\prime}(z)}\exp(\pm rW(z)),\label{eq:2}
\end{align}
where $r$ is a positive constant, $W$ is a monotonically increasing
function (i.e., $W^{\prime}>0$), and prime (${}^{\prime}$) indicates
the derivative with respect to $z$. In what follows we shall assume
that $W$ fulfills the following boundary conditions:
\begin{align}
  \lim_{z\to z_{\pm}}W(z)=\pm\frac{1}{2}.\label{eq:3}
\end{align}
$N$ is a normalization factor and chosen to satisfy
$\|\psi_{\pm}\|_{L^{2}(I)}=1$, where $\|\cdot\|_{L^{2}(I)}$ stands for
the $L^{2}$-norm on $I$.\footnote{The $L^{2}$-norm is defined by
  $\|f\|_{L^{2}(I)}=\sqrt{(f,f)_{L^{2}(I)}}$, where
  $(\cdot,\cdot)_{L^{2}(I)}$ stands for the $L^{2}$-inner product
  given by
  $(f,g)_{L^{2}(I)}=\int_{z_{-}}^{z_{+}}\!dz\,f^{\ast}(z)g(z)$ for any
  $f,g\in L^{2}(I)$.} It is easy to see that
$\|\psi_{\pm}\|_{L^{2}(I)}^{2}$ can be calculated without specifying
the explicit form of $W$ and take the following form:
\begin{align}
  \|\psi_{\pm}\|_{L^{2}(I)}^{2}
  &=|N|^{2}\int_{z_{-}}^{z_{+}}\!\!\!dz\,W^{\prime}(z)\exp\left(\pm2rW(z)\right)\nonumber\\
  &=|N|^{2}\left[\pm\frac{1}{2r}\exp\left(\pm2rW(z)\right)\right]_{z=z_{-}}^{z=z_{+}}\nonumber\\
  &=|N|^{2}\frac{\sinh(r)}{r},\label{eq:4}
\end{align}
where the last line follows from the boundary conditions
\eqref{eq:3}. Hence without any loss of generality $N$ can be chosen
as follows:
\begin{align}
  N=\sqrt{\frac{r}{\sinh(r)}}.\label{eq:5}
\end{align}

There are two important points to be emphasized here. The first is
that, thanks to the monotonicity of $W$, both $\psi_{+}$ and
$\psi_{-}$ are positive definite and have no node on $I$. Hence they
are good candidates for ground-state wavefunctions of one-dimensional
quantum-mechanical systems. The second is that, just like
Eq.~\eqref{eq:4}, the $L^{2}$-inner products
$(\psi_{\pm},\psi_{\mp})_{L^{2}(I)}$ and
$(\psi_{\pm},W\psi_{\pm})_{L^{2}(I)}$ are given by integrals of total
derivatives such that they can be calculated only from the boundary
conditions \eqref{eq:3}. In fact, a straightforward calculation gives
\begin{subequations}
  \begin{align}
    (\psi_{\pm},\psi_{\mp})_{L^{2}(I)}
    &=|N|^{2}\int_{z_{-}}^{z_{+}}\!\!\!dz\,W^{\prime}(z)\nonumber\\
    &=|N|^{2}\bigl[W(z)\bigr]_{z=z_{-}}^{z=z_{+}}\nonumber\\
    &=|N|^{2}\nonumber\\
    &=\frac{r}{\sinh(r)},\label{eq:6a}\\
    (\psi_{\pm},W\psi_{\pm})_{L^{2}(I)}
    &=|N|^{2}\int_{z_{-}}^{z_{+}}\!\!\!dz\,W(z)W^{\prime}(z)\exp\left(\pm2rW(z)\right)\nonumber\\
    &=|N|^{2}\left[\pm\frac{1}{2r^{2}}\left(rW(z)\mp\frac{1}{2}\right)\exp\left(\pm2rW(z)\right)\right]_{z=z_{-}}^{z=z_{+}}\nonumber\\
    &=|N|^{2}\left(\pm\frac{1}{2r}\cosh(r)\mp\frac{1}{2r^{2}}\sinh(r)\right)\nonumber\\
    &=\pm\frac{1}{2}\left(\coth(r)-\frac{1}{r}\right).\label{eq:6b}
  \end{align}
\end{subequations}
As we shall see shortly, these determine the BPS solutions of $SU(2)$
Yang-Mills-Higgs theory.

Now we wish to construct two distinct nodeless orthonormal
wavefunctions in order to fabricate doubly-degenerate ground
states. Eq.~\eqref{eq:6a}, however, implies that $\psi_{+}$ and
$\psi_{-}$ cannot be orthogonal with respect to the $L^{2}$-inner
product. However, they can become orthogonal if uplifted to the
vector-valued wavefunctions
$\left(\begin{smallmatrix}\psi_{+}\\0\end{smallmatrix}\right)$ and
$\left(\begin{smallmatrix}0\\\psi_{-}\end{smallmatrix}\right)$. More
generally, if we consider the two-component wavefunctions
\begin{align}
  \bm{\Psi}_{\pm}(z)=\psi_{\pm}(z)\bm{e}_{\pm},\label{eq:7}
\end{align}
where $\bm{e}_{+}$ and $\bm{e}_{-}$ stand for generic two-component
orthonormal complex vectors, $\bm{\Psi}_{+}$ and $\bm{\Psi}_{-}$
become orthonormal with respect to the inner product of the following
tensor-product Hilbert space:\footnote{The inner product on
  $\mathcal{H}=L^{2}(I)\otimes\mathbb{C}^{2}$ is defined by
  $(\bm{\Psi},\bm{\Phi})_{\mathcal{H}}=\int_{z_{-}}^{z_{+}}\!dz\,\bm{\Psi}^{\dagger}(z)\bm{\Phi}(z)$
  for any $\bm{\Psi},\bm{\Phi}\in\mathcal{H}$.}
\begin{align}
  \mathcal{H}=L^{2}(I)\otimes\mathbb{C}^{2}.\label{eq:8}
\end{align}
It should be noted that the unit vectors $\bm{e}_{\pm}$ are generally
parameterized by three independent reals. For the following
discussions we shall use the following parameterization:
\begin{align}
  \bm{e}_{+}=\frac{1}{\sqrt{2r(r-x_{3})}}\begin{pmatrix}x_{1}-ix_{2}\\r-x_{3}\end{pmatrix}
  \quad\text{and}\quad
  \bm{e}_{-}=\frac{1}{\sqrt{2r(r+x_{3})}}\begin{pmatrix}-x_{1}+ix_{2}\\r+x_{3}\end{pmatrix},\label{eq:9}
\end{align}
where $\bm{x}=(x_{1},x_{2},x_{3})\in\mathbb{R}^{3}\setminus\{0\}$ is a
nonvanishing 3-vector. It should be emphasized that in the above
parameterization we have identified the norm of $\bm{x}$ with the
parameter $r$ entering in the wavefunctions \eqref{eq:2}; that is,
$r=|\bm{x}|=\sqrt{(x_{1})^{2}+(x_{2})^{2}+(x_{3})^{2}}$. This
identification is technically essential in the following Berry
connection argument.

Now, let us suppose that there exists a quantum-mechanical system in
which the ground states are doubly degenerate and described by the
wavefunctions \eqref{eq:7}. Let us further assume that the three
parameters $\{x_{1},x_{2},x_{3}\}$ can be experimentally
controlled. Then, in such a system, under an adiabatic time-evolution
along a closed loop in the parameter space, the ground states acquire
a non-Abelian Berry phase described by the following Berry connection
\cite{Wilczek:1984dh}:
\begin{align}
  A_{ab}=i(\bm{\Psi}_{a},d\bm{\Psi}_{b})_{\mathcal{H}},\label{eq:10}
\end{align}
where $a,b\in\{+,-\}$ and $d=dx_{i}\frac{\partial}{\partial x_{i}}$
stands for the exterior derivative in the parameter space.  In the
following we shall also consider the following matrix elements of $W$:
\begin{align}
  \Phi_{ab}=(\bm{\Psi}_{a},W\bm{\Psi}_{b})_{\mathcal{H}}.\label{eq:11}
\end{align}
It should be noted that these quantities behave as a gauge field and
an adjoint Higgs field of $SU(2)$ gauge theories. Indeed, under a
unitary change of the basis (i.e., gauge transformation)
\begin{align}
  \bm{\Psi}_{a}\mapsto\Tilde{\bm{\Psi}}_{a}=\bm{\Psi}_{a^{\prime}}g_{a^{\prime}a},\label{eq:12}
\end{align}
where $g=(g_{a^{\prime}a})$ is a $2\times2$ unitary matrix,
$A=(A_{ab})$ and $\Phi=(\Phi_{ab})$ transform as the connection and
the adjoint representation for the Lie group $SU(2)$, respectively:
\begin{subequations}
  \begin{align}
    A&\mapsto\Tilde{A}=g^{\dagger}Ag+ig^{\dagger}dg,\label{eq:13a}\\
    \Phi&\mapsto\Tilde{\Phi}=g^{\dagger}\Phi g.\label{eq:13b}
  \end{align}
\end{subequations}

Now we wish to find explicit forms of $A$ and $\Phi$. To this end, it
is convenient to move to the gauge given by
$g=\left(\begin{smallmatrix}\bm{e}_{+}^{\dagger}\\\bm{e}_{-}^{\dagger}\end{smallmatrix}\right)$.
In this gauge Eqs.~\eqref{eq:13a} and \eqref{eq:13b} turn out to be of
the following forms:\footnote{To derive Eqs.~\eqref{eq:14a} and
  \eqref{eq:14b}, one first has to calculate $A=(A_{ab})$ and
  $\Phi=(\Phi_{ab})$, which take the following forms:
  \begin{align}
    A=
    \begin{pmatrix}
      i\bm{e}_{+}^{\dagger}d\bm{e}_{+}&iK\bm{e}_{+}^{\dagger}d\bm{e}_{-}\\
      iK\bm{e}_{-}^{\dagger}d\bm{e}_{+}&i\bm{e}_{-}^{\dagger}d\bm{e}_{-}\\
    \end{pmatrix}
    \quad\text{and}\quad
    \Phi=
    \begin{pmatrix}
      H&0\\
      0&-H\\
    \end{pmatrix},\nonumber
  \end{align}
  where $K=\frac{r}{\sinh(r)}$ and
  $H=\frac{1}{2}(\coth(r)-\frac{1}{r})$. These equations follow from
  \eqref{eq:6a}, \eqref{eq:6b},
  $\bm{e}_{a}^{\dagger}\bm{e}_{b}=\delta_{ab}$, and the identities
  $(\psi_{\pm},\frac{\partial}{\partial
    x_{i}}\psi_{\pm})_{L^{2}(I)}=\frac{\partial\log N}{\partial
    x_{i}}(\psi_{\pm},\psi_{\pm})_{L^{2}(I)}\pm\frac{\partial
    r}{\partial x_{i}}(\psi_{\pm},W\psi_{\pm})_{L^{2}(I)}=0$. For the
  gauge transformation induced by
  $g=\left(\begin{smallmatrix}\bm{e}_{+}^{\dagger}\\\bm{e}_{-}^{\dagger}\end{smallmatrix}\right)$,
  we refer to Appendix A of \cite{Ohya:2015xya}.}
\begin{subequations}
  \begin{align}
    \Tilde{A}&=\left(1-\frac{r}{\sinh(r)}\right)\epsilon_{ijk}\frac{x_{i}}{r^{2}}\frac{\sigma_{j}}{2}dx_{k},\label{eq:14a}\\
    \Tilde{\Phi}&=\left(\coth(r)-\frac{1}{r}\right)\frac{x_{i}}{r}\frac{\sigma_{i}}{2},\label{eq:14b}
  \end{align}
\end{subequations}
which are nothing but the celebrated BPS solutions for the
four-dimensional $SU(2)$ Yang-Mills-Higgs theory
\cite{Prasad:1975kr,Bogomolny:1975de}.

To summarize, we have found that a three-parameter family of mutually
orthogonal wavefunctions $\{\bm{\Psi}_{+},\bm{\Psi}_{-}\}$ yields the
BPS monopole and the adjoint Higgs field as the Berry connection and
the matrix elements of $W$. Note that $W$ is arbitrary except for the
monotonicity ($W^{\prime}>0$) and the boundary conditions
\eqref{eq:3}. This arbitrariness opens up a possibility to simulate
the BPS solutions in a wide range of nonrelativistic
quantum-mechanical systems, because there are infinitely many options
for such monotonic function. In the next Section we shall construct a
family of Schr\"{o}dinger Hamiltonians whose lowest-energy eigenstates
are all described by Eq.~\eqref{eq:7}.

\section{Model construction and exotic supersymmetry}
\label{section:3}
Now we wish to construct a $2\times2$ matrix-valued Hamiltonian for a
spin-1/2 particle which realizes Eq.~\eqref{eq:7} as the ground-state
wavefunctions. In fact, this is very easy to carry out once we realize
ground states generally determine potential energies. Below we shall
first outline the Hamiltonian construction and then discuss an exotic
supersymmetry hidden behind the energy spectrum.

To start with, let us first introduce the following first-order
differential operators:
\begin{subequations}
  \begin{align}
    D_{1}^{\pm}&=\pm\frac{d}{dz}-\frac{d\log\psi_{+}}{dz}=\pm\frac{d}{dz}-\frac{1}{2}\frac{W^{\prime\prime}}{W^{\prime}}-rW^{\prime},\label{eq:15a}\\
    D_{2}^{\pm}&=\pm\frac{d}{dz}+\frac{d\log\psi_{-}}{dz}=\pm\frac{d}{dz}+\frac{1}{2}\frac{W^{\prime\prime}}{W^{\prime}}-rW^{\prime}.\label{eq:15b}
  \end{align}
\end{subequations}
By construction it is obvious that $\psi_{+}$ and $\psi_{-}$ are the
zero-modes of $D_{1}^{+}$ and $D_{2}^{-}$, respectively. In other
words, they satisfy the first-order differential equations
$D_{1}^{+}\psi_{+}=0$ and $D_{2}^{-}\psi_{-}=0$. It is also obvious
that $D_{i}^{+}$ and $D_{i}^{-}$ ($i=1,2$) are hermitian conjugate
with each other with respect to the $L^{2}$-inner product on
$I$. Hence the second-order differential operator
$H_{\text{diag}}=\operatorname{diag}(D_{1}^{-}D_{1}^{+},D_{2}^{+}D_{2}^{-})$,
which is hermitian with respect to the inner product on $\mathcal{H}$,
is non-negative and enjoys doubly-degenerate ground states given by
$\left(\begin{smallmatrix}\psi_{+}\\0\end{smallmatrix}\right)$ and
$\left(\begin{smallmatrix}0\\\psi_{-}\end{smallmatrix}\right)$ with
the energy eigenvalue $E=0$. The unitary-transformed operator
$H=UH_{\text{diag}}U^{\dagger}$ thus provides the desired Hamiltonian
whose ground states are described by \eqref{eq:7}, provided $U$ is
chosen to satisfy
$U\left(\begin{smallmatrix}1\\0\end{smallmatrix}\right)=\bm{e}_{+}$
and
$U\left(\begin{smallmatrix}0\\1\end{smallmatrix}\right)=\bm{e}_{-}$. Note
that such $U$ is easily constructed and given by
$U=(\bm{e}_{+},\bm{e}_{-})$.

Having outlined the Hamiltonian construction, we are now ready to find
out the explicit form of $H$. Substituting Eqs.~\eqref{eq:15a} and
\eqref{eq:15b} into $H_{\text{diag}}$ we first get the following
diagonal Hamiltonian:
\begin{align}
  H_{\text{diag}}=\left[-\frac{d^{2}}{dz^{2}}+\frac{1}{2}\frac{W^{\prime\prime\prime}}{W^{\prime}}-\frac{1}{4}\left(\frac{W^{\prime\prime}}{W^{\prime}}\right)^{2}+r^{2}\left(W^{\prime}\right)^{2}\right]\bm{1}+2rW^{\prime\prime}\sigma_{3},\label{eq:16}
\end{align}
where $\bm{1}$ stands for the $2\times2$ identity matrix. Next, by
making use of the unitary transformation
\begin{align}
  H_{\text{diag}}\mapsto H=UH_{\text{diag}}U^{\dagger},\label{eq:17}
\end{align}
we finally get the following Hamiltonian for a spin-1/2
particle:\footnote{$\bm{\sigma}=(\sigma_{1},\sigma_{2},\sigma_{3})$.}
\begin{align}
  H=\left[-\frac{d^{2}}{dz^{2}}+\frac{1}{2}\frac{W^{\prime\prime\prime}}{W^{\prime}}-\frac{1}{4}\left(\frac{W^{\prime\prime}}{W^{\prime}}\right)^{2}+r^{2}\left(W^{\prime}\right)^{2}\right]\bm{1}+2W^{\prime\prime}\bm{x}\cdot\bm{\sigma},\label{eq:18}
\end{align}
where we have used
$U\sigma_{3}U^{\dagger}=\bm{e}_{+}\bm{e}_{+}^{\dagger}-\bm{e}_{-}\bm{e}_{-}^{\dagger}=(\bm{x}\cdot\bm{\sigma})/r$. This
is the Hamiltonian whose lowest-energy eigenstates are given by
\eqref{eq:7}. Note that the last term in Eq.~\eqref{eq:18} corresponds
to the interaction between the magnetic moment
$\bm{\mu}\propto\bm{\sigma}$ for a spin-1/2 particle and a
position-dependent external magnetic field
$\bm{B}(z)\propto W^{\prime\prime}(z)\bm{x}$. Note also that, if the
parameters $\bm{x}=(x_{1},x_{2},x_{3})$ are time-dependent and
adiabatically driven along a closed trajectory in the parameter space,
the doubly-degenerate ground states always acquire a non-Abelian Berry
phase described by the BPS monopole.

Now, Eq.~\eqref{eq:18} produces a large number of nonrelativistic
quantum-mechanical systems for a spin-1/2 particle by specifying the
subspace $I\subseteq\mathbb{R}$ and the monotonically increasing
function $W$. Before doing this, however, let us briefly point out
that our model possesses a hidden exotic supersymmetry called the
second-order derivative supersymmetry\footnote{In fact, the
  wavefunctions \eqref{eq:2} themselves have been obtained through the
  second-order derivative supersymmetry \cite{Andrianov:1994aj}.}
\cite{Andrianov:1993md,Andrianov:1994aj}, which is a nonlinear
extension of ordinary $\mathcal{N}=2$ supersymmetry. Its algebra
consists of four operators---the Hamiltonian $H$, the supercharge
$Q^{+}$ and its adjoint $Q^{-}=(Q^{+})^{\dagger}$, and the fermion
parity $(-1)^{F}$, the first three of which are second-order
derivative operators---and characterized by some nonlinear relation
among $H$ and $Q^{\pm}$. To see this, let us for the moment work in
the basis where the Hamiltonian becomes diagonal. Then, it is
straightforward to show that the $2\times2$ matrix-valued operators
\begin{subequations}
  \begin{align}
    H_{\text{diag}}&=\begin{pmatrix}D_{1}^{-}D_{1}^{+}&0\\0&D_{2}^{+}D_{2}^{-}\end{pmatrix},\label{eq:19a}\\
    Q^{+}&=\begin{pmatrix}0&0\\D_{2}^{+}D_{1}^{+}&0\end{pmatrix},\label{eq:19b}\\
    Q^{-}&=\begin{pmatrix}0&D_{1}^{-}D_{2}^{-}\\0&0\end{pmatrix},\label{eq:19c}\\
    (-1)^{F}&=\begin{pmatrix}1&0\\0&-1\end{pmatrix},\label{eq:19d}
  \end{align}
\end{subequations}
satisfy the following algebraic relations of second-order derivative
supersymmetry:
\begin{subequations}
  \begin{align}
    \left(Q^{\pm}\right)^{2}&=0,\label{eq:20a}\\
    \left((-1)^{F}\right)^{2}&=\bm{1},\label{eq:20b}\\
    [H_{\text{diag}},Q^{\pm}]&=0,\label{eq:20c}\\
    [H_{\text{diag}},(-1)^{F}]&=0,\label{eq:20d}\\
    \{Q^{\pm},(-1)^{F}\}&=0,\label{eq:20e}\\
    \{Q^{+},Q^{-}\}&=H_{\text{diag}}^{2},\label{eq:20f}
  \end{align}
\end{subequations}
where in Eqs.~\eqref{eq:20c} and \eqref{eq:20f} we have used the
following identity:\footnote{A straightforward calculation shows that
  $D_{1}^{+}D_{1}^{-}=D_{2}^{-}D_{2}^{+}=-\frac{d^{2}}{dz^{2}}-\frac{1}{2}S(W)+r^{2}(W^{\prime})^{2}$,
  where $S(W)$ stands for the Schwarzian derivative of $W$ given by
  $S(W)=\frac{W^{\prime\prime\prime}}{W^{\prime}}-\frac{3}{2}(\frac{W^{\prime\prime}}{W^{\prime}})^{2}$.}
\begin{align}
  D_{1}^{+}D_{1}^{-}=D_{2}^{-}D_{2}^{+}.\label{eq:21}
\end{align}
Note that these algebraic relations are invariant under the unitary
transformation $\mathcal{O}\mapsto U\mathcal{O}U^{\dagger}$,
$\mathcal{O}\in\{H_{\text{diag}},Q^{+},Q^{-},(-1)^{F}\}$. Hence the
quantum-mechanical system described by \eqref{eq:18} also possesses
this exotic supersymmetry. One of the big consequences of this
supersymmetry is that, in addition to the ground states, any other
discrete energy levels (if they exist) are guaranteed to be doubly
degenerate.

\section{Examples}
\label{section:4}
Before closing this paper let us present several examples of $H$ by
specifying $I$ and $W$. Since there are infinitely many options, in
this Section we will limit ourselves to only four illustrative
examples.

As noted at the end of the Introduction, we will proceed to use
arbitrary dimensionless units for notational simplicity.

\paragraph{Example 1: Hyperbolic tangent.} Let us first take $I$ as
the infinite interval $I=(-\infty,\infty)$. A typical example of
monotonically increasing function on $I$ that satisfies the boundary
conditions $\lim_{z\to\pm\infty}W(z)=\pm1/2$ is the following
hyperbolic tangent:
\begin{align}
  W(z)=\frac{1}{2}\tanh(z),\quad z\in(-\infty,\infty).\label{eq:22}
\end{align}
In this case the Hamiltonian \eqref{eq:18} turns out to be of the
following form:
\begin{align}
  H=\left[-\frac{d^{2}}{dz^{2}}-\frac{2}{\cosh^{2}(z)}+1+\frac{r^{2}/4}{\cosh^{4}(z)}\right]\bm{1}-2\frac{\sinh(z)}{\cosh^{3}(z)}\bm{x}\cdot\bm{\sigma}.\label{eq:23}
\end{align}
It should be noted that, in the limit $r=|\bm{x}|\to0$, the potential
energy in \eqref{eq:23} reduces to the famous reflectionless potential
that admits only one discrete energy level at the energy eigenvalue
$E=0$. The nonvanishing parameter $r$ hence describes the deformation
of reflectionless potential while keeping the double degeneracy of the
ground states.

\paragraph{Example 2: Error function.} Another example of
monotonically increasing function on $I=(-\infty,\infty)$ is the
following error function:
\begin{align}
  W(z)=\frac{1}{2}\operatorname{erf}(z),\quad z\in(-\infty,\infty),\label{eq:24}
\end{align}
where
$\operatorname{erf}(z)=\frac{2}{\sqrt{\pi}}\int_{0}^{z}\!dt\operatorname{e}^{-t^{2}}$. It
then follows from
$W^{\prime}(z)=\frac{1}{\sqrt{\pi}}\operatorname{e}^{-z^{2}}$ that the
Hamiltonian \eqref{eq:18} takes the following form:
\begin{align}
  H=\left[-\frac{d^{2}}{dz^{2}}+z^{2}-1+\frac{r^{2}}{\pi}\operatorname{e}^{-2z^{2}}\right]\bm{1}-\frac{4}{\sqrt{\pi}}z\operatorname{e}^{-z^{2}}\bm{x}\cdot\bm{\sigma}.\label{eq:25}
\end{align}
Notice that the potential energy in \eqref{eq:25} reduces to the
harmonic potential in the limit $r\to0$. Note also that, in contrast
to the previous example, the potential energy blows up in the limit
$z\to\pm\infty$ such that it describes a confining potential. Hence
the parameter $r$ describes the deformation of harmonic potential
while keeping the particle confinement and the ground-state
degeneracy.

\paragraph{Example 3: Trigonometric function.} Let us next consider
the case where $I$ is the finite interval $I=[0,\pi]$. In this case
$W$ can be chosen as the following trigonometric function:
\begin{align}
  W(z)=-\frac{1}{2}\cos(z),\quad z\in[0,\pi].\label{eq:26}
\end{align}
Substituting this into Eq.~\eqref{eq:18} we arrive at the following
Hamiltonian:
\begin{align}
  H=\left[-\frac{d^{2}}{dz^{2}}-\frac{1/4}{\sin^{2}(z)}-\frac{1}{4}+\frac{r^{2}}{4}\sin^{2}(z)\right]\bm{1}+\cos(z)\,\bm{x}\cdot\bm{\sigma}.\label{eq:27}
\end{align}
We emphasize that this is nothing but the model analyzed in
Ref.~\cite{Sonner:2008be}, where the authors have studied a spin-1/2
particle on $\mathbb{S}^{2}$ in the presence of a position-dependent
magnetic field as well as a particular external potential. In fact,
under the similarity transformation
$H\mapsto\Tilde{H}=(\sin(z))^{-1/2}H(\sin(z))^{1/2}$,
Eq.~\eqref{eq:27} is cast into the Hamiltonian essentially equivalent
to that used in \cite{Sonner:2008be}:
\begin{align}
  \Tilde{H}=\left[-\Delta_{\mathbb{S}^{2}}+\frac{r^{2}}{4}\sin^{2}(z)\right]\bm{1}+\cos(z)\,\bm{x}\cdot\bm{\sigma}.\label{eq:28}
\end{align}
where
$\Delta_{\mathbb{S}^{2}}=\frac{1}{\sin(z)}\frac{d}{dz}\sin(z)\frac{d}{dz}$
is the spherical Laplacian for functions independent of the polar
angle $\phi$. Note that $z$ should be read as the azimuthal angle
$\theta$. An important lesson from this example is that the
one-dimensional Hamiltonian \eqref{eq:18} can also be realized in
higher-dimensional systems through the separation of variables.

\paragraph{Example 4: Linear function.} Let us finally consider the
finite interval $I=[-\tfrac{1}{2},\tfrac{1}{2}]$ and the following
linear function:
\begin{align}
  W(z)=z,\quad z\in[-\tfrac{1}{2},\tfrac{1}{2}].\label{eq:29}
\end{align}
Since the second and third derivatives of the linear function
vanishes, in this case the Hamiltonian \eqref{eq:18} just becomes the
free Hamiltonian (with constant term):
\begin{align}
  H=\left[-\frac{d^{2}}{dz^{2}}+r^{2}\right]\bm{1}.\label{eq:30}
\end{align}
One might therefore think that this non-interacting system could not
exhibit any non-trivial non-Abelian Berry phase because the parameter
$\bm{x}$ disappears from the Hamiltonian. This is, however, not the
case because the parameter $\bm{x}$ can do appear in the boundary
conditions for the wavefunctions. In fact, it is well-known that the
self-adjoint extension argument leads to the $U(2)$ family of boundary
conditions at each boundary $z=\pm1/2$; see, e.g.,
Ref.~\cite{Fulop:1999pf}. In particular, a special thing happens
\cite{Ohya:2015xya} if we constrain ourselves to the following $SU(2)$
subfamily of boundary conditions:
\begin{align}
  (\bm{1}+U)\bm{\Psi}^{\prime}-i(\bm{1}-U)\bm{\Psi}=\bm{0}
  \quad\text{at}\quad
  z=\pm\tfrac{1}{2},\label{eq:31}
\end{align}
where $U\in SU(2)$. Note that any $SU(2)$ matrix can be parameterized
as
$U=\operatorname{e}^{i\alpha}P_{+}+\operatorname{e}^{-i\alpha}P_{-}$,
where $\alpha\in[0,\pi]$ and
$P_{\pm}=(\bm{1}\pm\Hat{\bm{x}}\cdot\bm{\sigma})/2$ with
$\Hat{\bm{x}}=\bm{x}/r$ being the unit vector pointing in the
direction of $\bm{x}$. Any element of $\mathcal{H}$ is then decomposed
as $\bm{\Psi}=\psi_{+}\bm{e}_{+}+\psi_{-}\bm{e}_{-}$ and
Eq.~\eqref{eq:31} boils down to the following Robin boundary
conditions for the coefficient functions $\psi_{\pm}$:
\begin{align}
  \pm\psi_{\pm}^{\prime}-\tan\left(\frac{\alpha}{2}\right)\psi_{\pm}=0
  \quad\text{at}\quad
  z=\pm\tfrac{1}{2},\label{eq:32}
\end{align}
which follow from $P_{\pm}\bm{e}_{\pm}=\bm{e}_{\pm}$ and
$P_{\pm}\bm{e}_{\mp}=\bm{0}$. If we identify $r=\tan(\alpha/2)$, the
ground-state wavefunctions can be written as
$\bm{\Psi}_{\pm}(z)=\sqrt{\frac{r}{\sinh(r)}}\exp(\pm
rz)\bm{e}_{\pm}$, which are exactly the same forms as
Eq.~\eqref{eq:7}. Hence the free spin-1/2 particle on the interval
with the particular boundary conditions \eqref{eq:31} also enjoys the
BPS Berry connection. An important lesson from this example is that
the parameter $\bm{x}$ does not always appear as the interaction
between the magnetic moment and the external magnetic field.

\bibliographystyle{utphys}%
\bibliography{bibliography}%
\end{document}